\documentclass[conference]{IEEEtran}
\IEEEoverridecommandlockouts
\usepackage{cite}
\usepackage{amsmath,amssymb,amsfonts}
\usepackage{algorithmic}
\usepackage{graphicx}
\usepackage{textcomp}
\usepackage{xcolor}
\def\BibTeX{{\rm B\kern-.05em{\sc i\kern-.025em b}\kern-.08em
    T\kern-.1667em\lower.7ex\hbox{E}\kern-.125emX}}
\begin{document}

\title{Sensitivity Quantification for Distribution System State Estimation\\

\thanks{This research is part of the research program ‘ROBUST - Trustworthy AI-based Systems for Sustainable Growth’, (partly) financed by the Dutch Research Council (NWO) through the LTP program with number KICH3.LTP.20.006.}
}

\author{\IEEEauthorblockN{Betül Mamudi}
\IEEEauthorblockA{\textit{Intelligent Electrical Power Grids} \\
\textit{Delft University of Technology}\\
Delft, The Netherlands \\
b.mamudi@tudelft.nl}
\and
\IEEEauthorblockN{Jochen B. Stiasny}
\IEEEauthorblockA{\textit{Intelligent Electrical Power Grids} \\
\textit{Delft University of Technology}\\
Delft, The Netherlands \\
j.b.stiasny@tudelft.nl}
\and
\IEEEauthorblockN{Jochen L. Cremer}
\IEEEauthorblockA{\textit{Intelligent Electrical Power Grids} \\
\textit{Delft University of Technology}\\
Delft, The Netherlands \\
j.l.cremer@tudelft.nl}

}

\maketitle

\begin{abstract}
Pseudo-measurements are the dominant source of uncertainty in 
distribution system state estimation (DSSE), yet their 
distributional assumptions are treated as fixed inputs by 
existing uncertainty quantification methods. This paper 
investigates whether the uncertainty bounds assumed by 
weighted least squares (WLS)-based DSSE are sensitive to 
these distributional assumptions, and whether this sensitivity 
is quantifiable using the Fisher Information Matrix (FIM). We 
propose a diagnostic framework that compares the true 
Cram\'{e}r--Rao Bound (CRB) against the WLS-assumed CRB via 
a per-bus, per-scenario ratio, computed directly from 
the converged WLS solution. Pseudo-measurement distributions 
are varied across five types in 22  
variants matched at equal spread to isolate shape effects from 
variance. Experiments on the CIGRE MV network across 100 
operating scenarios yield three findings. 

First, heavy-tailed and skewed distributions show consistently that WLS systematically overstates its uncertainty bounds. Second, the degree of 
miscalibration varies across buses and operating scenarios, 
confirming that distributional sensitivity is not uniform. 
Third, the CRB ratio is structurally blind to mean-shift bias, exposing a fundamental 
limitation of variance-based uncertainty diagnostics. Together, 
these results confirm the hypothesis and show that the choice 
of pseudo-measurement distribution directly distorts the 
confidence limits under WLS-based assumptions, which must be explicitly accounted for in any uncertainty-aware DSSE method.
\end{abstract}


\section{Introduction}
Power system state estimation is the operational foundation for monitoring and control of electrical power networks \cite{schweppe_power_1970}.
While methods for transmission systems are mature due to high measurement redundancy and comparatively simple topologies, these approaches are not directly applicable to distribution systems, which are characterized by sparse measurements and heterogeneous network topologies \cite{primadianto2017review}. Moreover, the rapid integration of decentralized generation and electrified loads have made reliable distribution system state estimation (DSSE) both more critical and more difficult \cite{moreno_escobar_comprehensive_2021,majeed_butt_recent_2021}.

The central obstacle to DSSE is limited observability \cite{della_giustina_electrical_2014}. Since full measurement instrumentation is economically infeasible,  so-called pseudo-measurements, statistically imputed data at unobserved buses, are used to close this gap \cite{pau_bayesian_2022}. These pseudo-measurements are typically derived from generic assumptions or historical load profiles, hence introducing structural uncertainty to the state estimates. Recent reviews study the generation of pseudo-measurements \cite{afrasiabi_pseudomeasurement_2025, ju_review_2025}, but do not systematically study the uncertainty introduced by wrongly specified statistical distributions. It is difficult to quantify the uncertainty that propagates to the DSSE results. With variance far exceeding the measurement noise of observed buses, pseudo-measurements are typically the dominant source of uncertainty in DSSE under known topology \cite{clements_impact_2011}.

The conventional DSSE method, weighted least squares (WLS), is a maximum-likelihood estimator under the assumption that measurement errors are zero-mean Gaussian \cite{abur2004power}. 
This assumption is convenient but demonstrably wrong for pseudo-measurements, which exhibit heavy tails, asymmetry, and systematic bias \cite{pegoraro_bayesian_2017}.
Prior work has shown that non-Gaussian noise degrades WLS accuracy \cite{cubonovic_impact_2024} and that using the correct likelihood improves results \cite{vanin_exact_2023}, but neither asks the more fundamental diagnostic question: \emph{By how much do distributional assumptions distort the reported uncertainty?} As this paper demonstrates, the answer can exceed 
a factor of four in the variance lower bound.

Bayesian state estimation yields posterior uncertainty quantification (UQ) under an assumed prior, but its quality relies on the distributional assumptions under examination \cite{pegoraro_bayesian_2017}. For distribution system operators, this potentially miscalibrated uncertainty can lead to overconfident operational decisions or outages \cite{pau_impact_2019}. To the best of the authors' knowledge, no existing approaches for DSSE explicitly measure the cost of distributional miscalibration.

The Fisher Information Matrix (FIM) provides a computationally efficient framework to analyze estimator uncertainty \cite{kay_fundamentals_1993}. It quantifies how informative each measurement is with respect to the state variables. Its inverse, the Cramér-Rao Bound (CRB), gives a lower bound on the variance of an unbiased estimator. Under Gaussian assumptions, the WLS gain matrix equals the FIM \cite{abur2004power}. This shows that any miscalibrated pseudo-measurement distribution directly propagates to the reported uncertainty limits. When pseudo-measurement errors 
are non-Gaussian, the true scalar Fisher information deviates from the 
Gaussian weight assumed by WLS, creating a quantifiable mismatch between 
the uncertainty WLS claims and the minimum variance achievable with the true likelihood.

This paper tests the hypothesis: \emph{``The uncertainty bounds assumed by WLS-based DSSE are sensitive to the distributional assumptions of the pseudo-measurements, and this sensitivity is quantifiable using Fisher Information.''} 

We propose a systematic sensitivity analysis framework using the CRB as a diagnostic tool and varying pseudo-measurement distributions across five types (Gaussian, Student-t, Laplace, skew-normal, biased Gaussian) in 22 variants matched at equal spread to isolate shape effects from variance effects. The framework is estimator-agnostic. The goal is to make the cost of distributional miscalibration explicit and measurable.

\section{Distribution System State Estimation}
\label{sec:dsse_background}

In a distribution system with $n_b$ buses, the \emph{state vector}
$x \in \mathbb{R}^{n_s}$ with $n_s = 2(n_b - 1)$ collects the voltage phasors at all non-slack buses,
\begin{equation}
  {x} =
    \bigl[\theta_1, \ldots, \theta_{n_b-1},\;
          |V_1|, \ldots, |V_{n_b-1}|\bigr]^{\top},
  \label{eq:state_vec}
\end{equation}
where $\theta_i$ and $|V_i|$ denote the voltage angle (rad) and magnitude (p.u.) at bus $i$. Bus~0 is the slack reference with a fixed voltage magnitude $|V_0|$  and $\theta_0 = 0$. The state is related to the measurements through the nonlinear measurement model,
\begin{equation}
  {z} = \mathbf{h}({x}) + \boldsymbol{\varepsilon},
  \label{eq:meas_model_intro}
\end{equation}
where ${z} \in \mathbb{R}^{m}$ is the measurement vector, $\mathbf{h} : \mathbb{R}^{n_s} \to \mathbb{R}^{m}$ maps the state through the AC power-flow equations, and $\boldsymbol{\varepsilon}$ is the measurement error vector \cite{abur2004power}.

WLS minimizes the weighted sum of squared residuals,
\begin{equation}
  \hat{{x}}_{\mathrm{WLS}}
  = \operatorname*{arg\,min}_{{x}}\;
    \bigl[{z} - \mathbf{h}({x})\bigr]^{\top}
    \mathbf{W}
    \bigl[{z} - \mathbf{h}({x})\bigr],
  \label{eq:wls_intro}
\end{equation}
where $\mathbf{W} = \mathrm{diag}(1/\sigma_1^2, \ldots, 1/\sigma_m^2)$ under the Gaussian assumption $\varepsilon_i \sim \mathcal{N}(0, \sigma_i^2)$. Because $\mathbf{h}$ is nonlinear, \eqref{eq:wls_intro} is solved
iteratively via Newton--Raphson. At each iteration $k$, the \emph{normal equations}
\begin{equation}
  \mathbf{G}^{(k)}\,\Delta{x}^{(k)}
  = \mathbf{H}^{\top}({x}^{(k)})\,\mathbf{W}\,
    \bigl[{z} - \mathbf{h}({x}^{(k)})\bigr]
  \label{eq:normal_intro}
\end{equation}
are solved for the correction step $\Delta{x}^{(k)}$, where the Jacobian is
$\mathbf{H}({x}) = \partial\mathbf{h}/\partial{x}^{\top}
\in \mathbb{R}^{m \times n_s}$ and
\begin{equation}
  \mathbf{G}^{(k)}
  = \mathbf{H}^{\top}({x}^{(k)})\,\mathbf{W}\,
    \mathbf{H}({x}^{(k)})
  \label{eq:gain_intro}
\end{equation}
is the \emph{gain matrix}. The update ${x}^{(k+1)} =
{x}^{(k)} + \Delta{x}^{(k)}$ is applied until
$\|\Delta{x}^{(k)}\|_{\infty} < \varepsilon_{\mathrm{tol}}$,
where $\varepsilon_{\mathrm{tol}}$ is the convergence tolerance.
For an observable system, the Jacobian $\mathbf{H}({x})$ needs to have full column rank ($\mathrm{rank}(\mathbf{H})=n_s$), requiring the number of measurements $m \geq n_s$. 

Since distribution networks
are not fully observable, pseudo-measurements are introduced at unobserved buses to achieve observability \cite{pau_bayesian_2022}. 
Pseudo-measurements are modeled as random variables $z_i = \mu_i^*+\epsilon_i$, where $\mu_i^*$ is the true load at bus $i$ and $\epsilon_i \sim p_i(0, \sigma_i^2)$ is a nominally zero-mean noise term with spread $\sigma_i$, whose distributional assumptions are the subject of the sensitivity analysis. 

\section{Proposed Sensitivity Quantification}

\subsection{Fisher Information Matrix and Cramér-Rao Bound}
The FIM quantifies how much information the measurements ${z}$ carry about the state ${x}$. For independent measurements with log-likelihood $\ell_i({x}; z_i) = \log p(z_i \mid {x})$, the FIM is the negative expected Hessian of the total log-likelihood
\cite{kay_fundamentals_1993},
\begin{equation}
  \mathcal{I}(\mathbf{x})
  = -\mathbb{E}\!\left[
      \frac{\partial^2 \ell({x};{z})}
           {\partial {x}\,\partial {x}^{\top}}
    \right]
  = \mathbb{E}\!\left[
      \frac{\partial \ell}{\partial {x}}
      \frac{\partial \ell}{\partial {x}^{\top}}
    \right],
  \label{eq:fim}
\end{equation}
where the expectation is taken over ${z}$ given ${x}$.
The FIM equals the curvature of the log-likelihood at its maximum. Thus,
a sharper peak means more information and lower uncertainty in the estimation. Formally, this curvature interpretation follows from
the second-order Taylor expansion of $\ell$ around the
maximum-likelihood estimate $\hat{{x}}$ \cite{bishop_pattern_2006},
\begin{equation}
  \ell({x};{z})
  \approx \ell(\hat{{x}};{z})
  - \frac{1}{2}
    ({x} - \hat{{x}})^{\top}
    \mathcal{I}(\hat{{x}})
    ({x} - \hat{{x}}),
  \label{eq:laplace}
\end{equation}
which is the Laplace approximation. The posterior is approximated
as a Gaussian with covariance $\mathcal{I}^{-1}(\hat{{x}})$
\cite{kay_fundamentals_1993}.
This approximation is exact only when the likelihood is Gaussian and
the measurement model is linear \cite{bishop_pattern_2006}.

The CRB is the inverse of the FIM and states that the covariance for any unbiased estimator satisfies $\mathrm{Cov}(\hat{\mathbf{x}}) \succeq \mathcal{I}^{-1}(\mathbf{x})$, thus specifying the lower bound of the variance per state variable based on the measurement distribution \cite{kay_fundamentals_1993}.

\subsection{Connection to the WLS Gain Matrix}

The structural link between the FIM and WLS emerges when the
measurement errors are independent and Gaussian,
$\varepsilon_i \sim \mathcal{N}(0, \sigma_i^2)$. The log-likelihood
then takes the form
\begin{equation}
  \ell({x};{z})
  = -\frac{1}{2}\sum_{i=1}^{m}
    \frac{[z_i - h_i({x})]^2}{\sigma_i^2}
    + \text{const.},
  \label{eq:gaussian_ll}
\end{equation}
which is precisely the negative of the WLS objective in
\eqref{eq:wls_intro}. Substituting \eqref{eq:gaussian_ll} into
\eqref{eq:fim} and evaluating the linearized measurement
model $\mathbf{h}({x}) \approx \mathbf{H}{x}$
yields
\begin{equation}
  \mathcal{I}({x})
  = \mathbf{H}^{\top}\mathbf{W}\mathbf{H}
  = \mathbf{G},
  \label{eq:fim_equals_gain}
\end{equation}
where $\mathbf{W} = \mathrm{diag}(1/\sigma_1^2, \ldots,
1/\sigma_m^2)$ \cite{abur2004power}. Equation~\eqref{eq:fim_equals_gain}
shows that under Gaussian noise and a linear measurement model,
the FIM is exactly the WLS gain matrix $\mathbf{G}$, and its
inverse $\mathbf{G}^{-1}$ is the CRB. When either condition fails, the equality in \eqref{eq:fim_equals_gain} does not hold and $\mathbf{G}^{-1}$ no longer correctly represents the true estimation uncertainty.

\subsection{True and Assumed Variance}

When the true noise distribution $p_i(\varepsilon_i)$ is non-Gaussian, the scalar Fisher information of measurement $i$ is 
\begin{equation}
  \mathcal{F}_i
  = \int_{-\infty}^{\infty}
    \!\left[\frac{d}{d\varepsilon}
      \log p_i(\varepsilon)\right]^{\!2}
    p_i(\varepsilon)\,d\varepsilon,
  \label{eq:scalar_fi}
\end{equation}
which equals $1/\sigma_i^2$ only for Gaussian distributions \cite{kay_fundamentals_1993}. This
motivates defining two distinct gain matrices at the converged
estimate $\hat{{x}}$,
\begin{equation}
  \mathbf{G}_{\mathrm{WLS}}
  = \mathbf{H}^{\top}\mathbf{W}_{\mathrm{assumed}}\mathbf{H},
  \qquad
  \mathbf{G}_{\mathrm{true}}
  = \mathbf{H}^{\top}\mathbf{W}_{\mathrm{true}}\mathbf{H},
  \label{eq:two_gains}
\end{equation}
where $\mathbf{W}_{\mathrm{assumed}} =
\mathrm{diag}(1/\sigma_{\mathrm{assumed},i}^2)$ uses the variance
of the Gaussian distribution, and $\mathbf{W}_{\mathrm{true}} =
\mathrm{diag}(\mathcal{F}_1, \ldots, \mathcal{F}_m)$ uses the
true scalar Fisher information values from \eqref{eq:scalar_fi}.
The matrix $\mathbf{G}_{\mathrm{WLS}}^{-1}$ is the uncertainty
the estimator \emph{claims}; $\mathbf{G}_{\mathrm{true}}^{-1}$ is
the true CRB. Their element-wise ratio,
\begin{equation}
  \rho_k
  = \frac{[\mathbf{G}_{\mathrm{true}}^{-1}]_{kk}}
         {[\mathbf{G}_{\mathrm{WLS}}^{-1}]_{kk}},
  \label{eq:crb_ratio}
\end{equation}
is the \emph{CRB ratio} for state variable $k$. 
The Gaussian baseline recovers $\rho_k = 1$, since assumed
and true Fisher information coincide. When $\rho_k < 1$, the
true CRB is tighter than what WLS claims: the non-Gaussian
distribution carries more Fisher information than the Gaussian
assumption credits, WLS overstates its uncertainty bounds.
When $\rho_k > 1$, the distribution is less informative and WLS underestimates its uncertainty. 

Computing \eqref{eq:crb_ratio} across different statistical distribution types and operating scenarios makes the sensitivity of WLS uncertainty bounds directly quantifiable.

\section{Sensitivity Analysis on CIGRE MV Network}

\subsection{Case Study Setup}
\begin{figure}[b]
    \centering
    \includegraphics[width=0.45\linewidth]{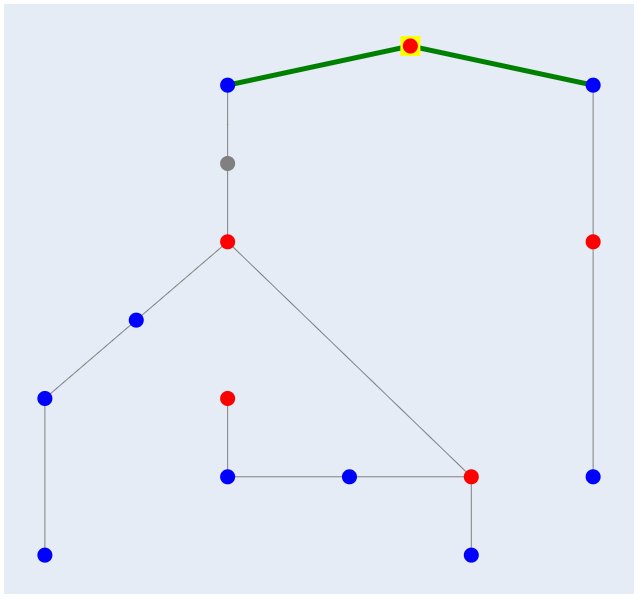}
    \caption{CIGRE MV Network. Slack bus (yellow); buses with real measurements (red); zero-injection bus 2 (grey); measured feeder heads (green).}
    \label{fig:cigre}
\end{figure}

The CIGRE MV network is used as the test system, instantiated in pandapower. The network comprises $n_b = 15$ buses operating at 20\,kV, giving a state vector of dimension $n_s = 28$. Two normally-open switches are excluded from the admittance matrix, leaving a radial operating configuration with two feeders, as shown in Fig. \ref{fig:cigre}.

100 independent operating scenarios are generated by
sampling a load scaling factor $\lambda \sim \mathcal{U}(0.5, 1.5)$
independently for each scenario and applying it uniformly to the
active and reactive power at every load bus, preserving
the per-bus power factor. For each scenario, an AC power flow is
solved via Newton--Raphson with convergence tolerance
$10^{-8}$\,MVA; non-converging scenarios are discarded and resampled. The converged bus voltages constitute the true state ${x}^*$, which
serves as ground truth for the evaluation later on.

The measurement vector ${z} \in \mathbb{R}^{36}$ combines
real measurements and pseudo-measurements
at unobserved buses. Real measurements consist of voltage
magnitudes at buses 0, 3, 8, 11, 13; active and reactive
power injections at buses 3, 8, 11, 13; and active and
reactive line flows at feeder heads.
These 17 measurements are corrupted by additive Gaussian noise,
\begin{equation}
  z_i = z_i^* + \varepsilon_i, \quad
  \varepsilon_i \sim \mathcal{N}(0,\, \sigma_i^2), \quad
  \sigma_i = \sigma_{\%,i} \cdot |z_i^*|,
  \label{eq:noise}
\end{equation}
where $\sigma_{\%,i}$ is drawn once per bus from
$\mathcal{U}(0.5\%, 2\%)$ for voltages and
$\mathcal{U}(1\%, 5\%)$ for loads, then fixed
across all 100 scenarios to reflect sensor accuracy.

Pseudo-measurements are assigned to the 10 unobserved buses, providing active and reactive power injections, giving 20 pseudo-measurements per loading scenario. Bus 2 is a zero-injection bus. Its pseudo-measurement
is assigned a variance of $\sigma^2_{\text{floor}} =
10^{-4}$\,p.u.$^2$ to enforce the zero-injection constraint without modifying the WLS formulation.

To systematically study the sensitivity of WLS uncertainty to
distributional shape, pseudo-measurements are sampled
from five distribution types across 22 variants, all matched at equal spread $\sigma_i = \sigma_\% \cdot |\mu_i^*|$, where $\mu_i^*$ is the true load at bus $i$ (see
Table~\ref{tab:variants}).

\begin{table}[t]
  \renewcommand{\arraystretch}{1.2}
  \caption{Pseudo-Measurement Distribution Variants}
  \label{tab:variants}
  \centering
  \begin{tabular}{llr}

    Type & Parameter & Variants \\
    \hline
    Gaussian      & $\sigma_\% \in \{10, 20, 30\}\%$         & 3 \\
    Student-$t$   & $\sigma_\% \times \nu \in \{3,4\}$        & 6 \\
    Laplace       & $\sigma_\% \in \{10, 20, 30\}\%$         & 3 \\
    Skew-normal   & $\sigma_= 20\%; \alpha \in \{2, 5, 7, 10\}$             & 4 \\
    Biased Gaussian & $\sigma_= 20\%$; bias $\in \{{\pm}10, {\pm}20, {\pm}30\}\%$ & 6 \\

    Total         &                                           & 22 \\

  \end{tabular}
\end{table}

The families are chosen to represent the empirically relevant inaccuracies of the Gaussian assumption in real load data. \emph{Gaussian} 
serves as the calibrated baseline, for which $\rho_k = 1$ by 
construction; \emph{Student-t} captures heavy tails common in aggregated consumption profiles, where rare spikes inflate tail probability; \emph{Laplace} is the maximum-entropy distribution at fixed mean absolute deviation and represents sharp-peaked, heavy-tailed noise; \emph{Skew-normal} is mode-centered on $\mu_i^*$, reflecting that real load profiles peak at the typical operating point while occasional demand spikes produce a heavier right tail; \emph{Biased Gaussian} models systematically wrong prior 
information, such as incorrect customer classification, by shifting the mean by a fixed fraction of $\mu_i^*$ while keeping the Gaussian.

For the Student-$t$ family, the scale parameter is adjusted
to $\sigma_i / \sqrt{\nu / (\nu - 2)}$ so that the marginal
standard deviation matches $\sigma_i$ for $\nu \in \{3, 4\}$,
enabling a fair comparison with the other families at equal
spread. For the Laplace family, the scale parameter is set to
$b_i = \sigma_i / \sqrt{2}$, which yields a variance of
$\sigma_i^2$, again matching the spread of the other families. The skew-normal distribution is right-skewed for $\alpha > 0$. This makes $\mathrm{E}[X] > \mu_i^*$, introducing a mean shift as a structural consequence of asymmetry rather than an
independent bias parameter. The biased Gaussian instead
shifts the mean by a fixed fraction of $\mu_i^*$ while
keeping the spread at $\sigma_\% = 20\%$.

\subsection{Sensitivity to Distributional Assumptions}

Three complementary analyses are conducted. The CRB ratio quantifies miscalibration in the uncertainty bounds based on the Fisher information. Empirical coverage validates whether this miscalibration is present in realized estimation error. Root Mean Square Error (RMSE) finally checks whether distributional shape also affects point accuracy.

\subsubsection{CRB ratio}
\begin{figure}[t]
    \centering
    \includegraphics[width=0.9\linewidth]{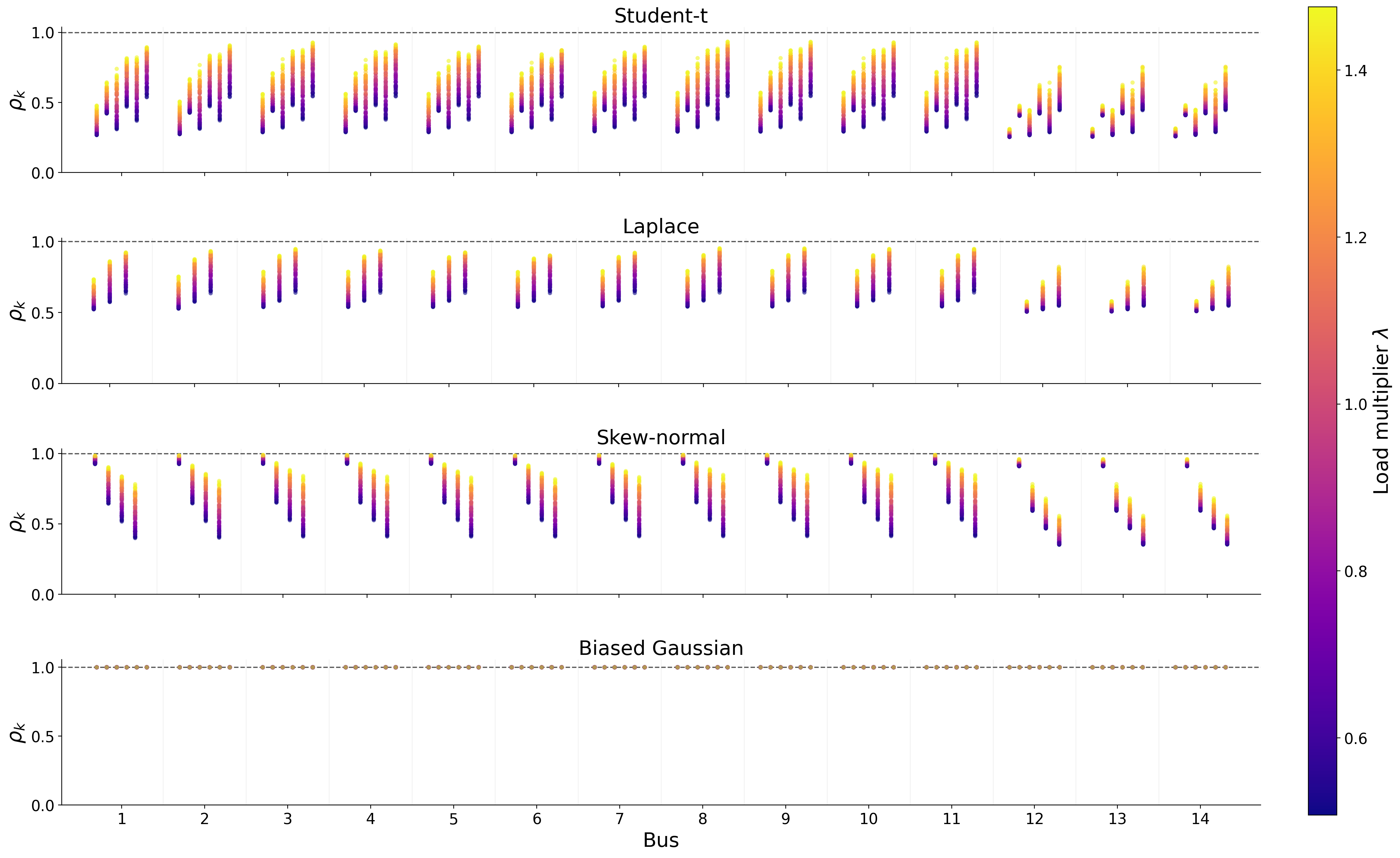}
    \caption{CRB Ratio $\rho_k$ for voltage magnitude at all 14 non-slack buses (x-axis). 100 operating scenarios (dots). Results for all distribution variants (see Table \ref{tab:variants}). $\rho_k<1$ indicates that WLS overstates the uncertainty bounds.}
    \label{fig:crb_ratio}
\end{figure}

Fig.~\ref{fig:crb_ratio} shows the CRB ratio $\rho_k$ for
voltage magnitudes across all 14 non-slack buses and 100 scenarios.

For Student-$t$, Laplace, and Skew-normal, $\rho_k < 1$ consistently 
across all buses and scenarios. These distributions carry more Fisher 
information than the Gaussian assumption shows, therefore WLS overstates 
its uncertainty bounds. The degree of miscalibration differs 
substantially by family. 

Student-$t$ produces the most severe 
miscalibration with the minimum observed ratio $\rho_k = 0.255$ 
(bus~12, $\nu=3$, $\sigma_\%=10\%$, $\lambda=0.507$), meaning WLS 
overstates the variance bound by nearly a factor of four, and 
92.2\% of all bus--scenario pairs for this distribution variant fall below 
$\rho_k = 0.5$. Across all Student-$t$ distributions, 42.7\% of the ratios lie 
below this threshold. Laplace distributions, by contrast, remain entirely above 
$\rho_k = 0.5$ across all variants (minimum 0.506). The heavy tails cause miscalibration, but the double-exponential decay is less severe than the polynomial 
tails of Student-$t$. Skew-normal lies between, with 10.7\% 
of pairs below $\rho_k = 0.5$ and $\rho_k$ approaching 1 at low skewness 
($\alpha=2$), indicating that mild asymmetry alone causes less variance-based
miscalibration than heavy tails.
The Biased Gaussian shows $\rho_k = 1$ for
all buses, scenarios, and bias levels. This confirms that a mean
shift in the pseudo-measurement distribution does not affect
the Fisher information and therefore leaves the CRB ratio
unchanged. The CRB ratio is thus blind to systematic mean error in
pseudo-measurements.

A spatial structure is visible across all three families.
Buses on the shorter feeder, particularly buses 12--14,
exhibit lower $\rho_k$ and wider inter-scenario spread
than buses on the main feeder. 
The load multiplier $\lambda$ shows a moderate influence. The worst-case operating point is consistently the lightest loading ($\lambda \approx 0.507$) and higher loading scenarios (yellow) tend toward larger $\rho_k$. This indicates an operating-point and spatial dependence of the CRB sensitivity.

\subsubsection{Empirical Coverage}
To validate the CRB ratio against realized estimation error, we compute empirical coverage, the fraction of scenarios in which the true state falls within the interval implied by $\mathbf{G}^{-1}$ at nominal level $\alpha$: 
\begin{equation}
  \widehat{\mathrm{Cov}}_{\alpha,k}
  = \frac{1}{N}\sum_{c=1}^{N}
    \mathbf{1}\!\left[
      |x_k^* - \hat{x}_k^{(c)}|
      < z_\alpha \sqrt{[\mathbf{G}^{-1}]_{kk}}
    \right],
  \label{eq:coverage}
\end{equation}
where $N = 1000$ is the number of scenarios, $z_{0.68} = 1.0$
and $z_{0.95} = 1.96$ are the Gaussian quantiles, and
$\mathbf{G}^{-1}$ is either $\mathbf{G}_{\mathrm{WLS}}^{-1}$
(assumed) or $\mathbf{G}_{\mathrm{true}}^{-1}$. 
A larger $N$ is used here than for the CRB ratio analysis to reduce noise at the 95\% nominal level. Values are averaged over all 14 non-slack buses. A perfectly calibrated estimator achieves
$\widehat{\mathrm{Cov}}_\alpha = \alpha$.

Table~\ref{tab:coverage} reports these values for all 22 variants. For the Gaussian family, assumed and true coverage coincide and remain close to the nominal levels, confirming that WLS is well-calibrated under its own assumption.
For Student-$t$ and Laplace, the assumed coverage at the $\pm\sigma_\%$--interval
exceeds the nominal level (ranging from 0.70 to 0.77), consistent with the $\rho_k < 1$ finding,
while the true coverage falls below it (0.52--0.64).
The Skew-normal shows a similar pattern with increasing severity as $\alpha$
grows.
For the Biased Gaussian, assumed and true coverage coincide, consistent with $\rho_k = 1$, but both fall
significantly below the nominal level at large bias
magnitudes, reaching 0.30 at 68\% for a $- 30\%$ bias. This
reveals the limitation identified earlier. At small biases the wide pseudo-measurement variance ($\sigma_\% = 20\%$) partially absorbs the shift, maintaining near-nominal coverage; at larger biases the systematic offset moves the estimate far enough from truth that the wide interval no longer covers it. The CRB
ratio cannot detect this degradation, since it captures only the curvature of the log-likelihood and not its location.
\begin{table}[b]
\caption{Empirical Coverage at 68\% and 95\% Nominal Levels,
Averaged over all non-slack buses and 1000 loading scenarios.}
\label{tab:coverage}
\centering
\begin{tabular}{ll|cccc}
Type & Parameter
  & $\mathrm{Cov}_{68}^{\mathrm{WLS}}$
  & $\mathrm{Cov}_{68}^{\mathrm{true}}$
  & $\mathrm{Cov}_{95}^{\mathrm{WLS}}$
  & $\mathrm{Cov}_{95}^{\mathrm{true}}$ \\
\hline
Gaussian        & 10\% & 0.702 & 0.702 & 0.944 & 0.944 \\
( $\sigma_\%$)                & 20\% & 0.665 & 0.665 & 0.943 & 0.943 \\
                & 30\% & 0.693 & 0.693 & 0.952 & 0.952 \\
\hline
Student-$t$     & (10\%,3) & 0.768 & 0.516 & 0.952 & 0.819 \\
$(\sigma_\%,\nu)$               & (10\%,4) & 0.740 & 0.575 & 0.956 & 0.864 \\
                & (20\%,3) & 0.738 & 0.529 & 0.950 & 0.826 \\
                & (20\%,4) & 0.730 & 0.584 & 0.960 & 0.881 \\
                & (30\%,3) & 0.741 & 0.547 & 0.962 & 0.855 \\
                & (30\%,4) & 0.728 & 0.609 & 0.961 & 0.891 \\
\hline
Laplace         & 10\% & 0.708 & 0.590 & 0.946 & 0.869 \\
($\sigma_\%$)                & 20\% & 0.718 & 0.615 & 0.951 & 0.890 \\
                & 30\% & 0.706 & 0.635 & 0.950 & 0.911 \\

\hline
Skew-     &  2   & 0.673 & 0.658 & 0.950 & 0.941 \\
normal                &  5   & 0.606 & 0.529 & 0.904 & 0.835 \\
($\alpha$)                &  7   & 0.576 & 0.466 & 0.884 & 0.773 \\
                & 10   & 0.546 & 0.399 & 0.851 & 0.689 \\

\hline
Biased          & $-30\%$ & 0.304 & 0.304 & 0.653 & 0.653 \\
Gaussian        & $-20\%$ & 0.474 & 0.474 & 0.805 & 0.805 \\
(bias)                & $-10\%$ & 0.613 & 0.613 & 0.913 & 0.913 \\
                & $+10\%$ & 0.607 & 0.607 & 0.923 & 0.923 \\
                & $+20\%$ & 0.487 & 0.487 & 0.835 & 0.835 \\
                & $+30\%$ & 0.315 & 0.315 & 0.691 & 0.691 \\

\end{tabular}
\end{table}
\subsubsection{WLS Estimation Accuracy}
\begin{figure}[t]
    \centering
    \includegraphics[width=0.8\linewidth]{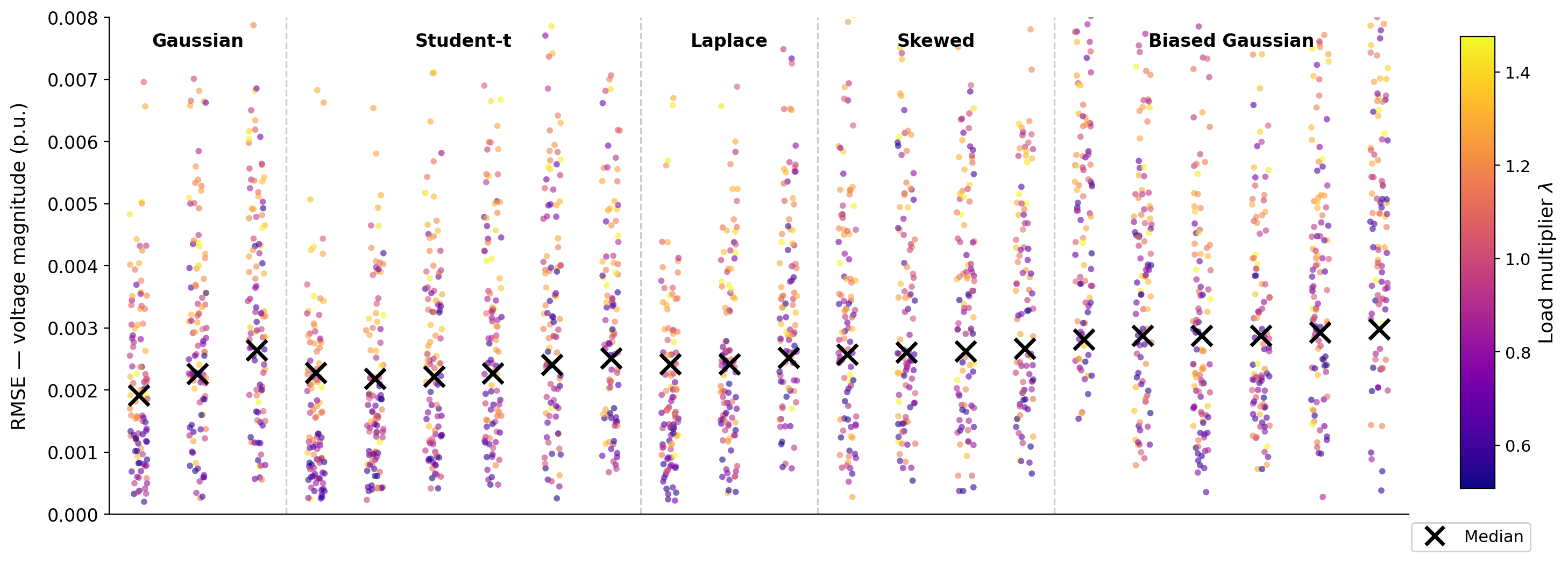}
    \caption{RMSE of WLS-based DSSE under distributional assumptions}
    \label{fig:rmse}
\end{figure}

Fig. \ref{fig:rmse} shows the RMSE on voltage magnitudes across all 22 distribution variants, coloured by load multiplier $\lambda$. The Gaussian distributions show that median RMSE increases with $\sigma \%$. Across non-Gaussian distributions, median RMSE remains broadly comparable to the Gaussian baseline at equal $\sigma \%$. Consequently, the distributional shape affects the estimator's uncertainty bounds more than point accuracy when the distribution is correctly centered.

The clearest deviation appears in the Biased Gaussian variants, where the median RMSE is elevated and increases with bias magnitude, despite $\rho_k=1$. This confirms from a third angle the limitation identified above: a systematic mean shift degrades estimation accuracy in a way that is invisible to variance-based diagnostics but directly observable in empirical error. The $\lambda$-dependence visible across all variants reflects the proportional noise model $\sigma_i=\sigma\% \cdot  |\mu^*_i|$.

\section{Conclusion}
This paper investigated the hypothesis that WLS-based DSSE uncertainty bounds are sensitive to pseudo-measurement distributional assumptions, and that this sensitivity is quantifiable using Fisher Information. Experiments on the CIGRE MV network across 100 operating scenarios and 22 distribution variants confirm this hypothesis with three findings:

\begin{enumerate}

    \item For Student-$t$, Laplace, and Skew-normal pseudo-measurement distributions, $\rho_k < 1$ across all buses and scenarios. WLS systematically overstates its uncertainty bounds because these distributions carry more Fisher information than the Gaussian assumption credits.

    \item The miscalibration varies across buses and load scenarios, implying that the sensitivity depends on the network structure and load conditions. It is not a global scalar, but a spatially and operationally structured quantity.

    \item The CRB ratio is structurally blind to mean shift: a biased distribution leaves $\rho_k = 1$ while degrading empirical coverage to 0.30 at the 68\% nominal level. This exposes a fundamental limitation of variance-based UQ diagnostics, which capture the curvature of the log-likelihood but not its location, and therefore cannot detect systematic bias regardless of its magnitude.

\end{enumerate}

These results show that the choice of
pseudo-measurement distribution is not a modelling detail
but a structural property of the estimation problem that
directly distorts the confidence limits reported by WLS. This should be explicitly accounted for in any
uncertainty-aware DSSE method. Future work
should investigate coverage-based alternatives such as
conformal prediction as a complementary diagnostic that
captures location shift alongside curvature, and extend the
framework to larger, more realistic distribution networks.

\section*{Acknowledgment}
B.M. would like to thank Werner van Westering and Jacco Heres for their helpful guidance regarding the study’s configuration.
The authors have declared no conflicts of interest.
The generation of this manuscript was assisted by Claude
(Anthropic), a generative AI tool, for drafting and editing
text, in accordance with IEEE guidelines on AI-assisted
writing.

\bibliographystyle{IEEEtran}
\bibliography{references}

\end{document}